\documentclass[epj]{svjour}

\usepackage{graphics}
\usepackage{amssymb}

\begin{document}

\title{Enhancement of antiferromagnetic interaction and transition temperature in $M_3$TeO$_6$ systems ($M$ = Mn, Co, Ni, Cu)}

\author{R. Mathieu\inst{1}, S. A. Ivanov\inst{1,2}, P. Nordblad\inst{1}, and M. Weil\inst{3}}

\authorrunning{R. Mathieu et al.}

\titlerunning{Enhancement of antiferromagnetic interaction and $T_N$ in $M_3$TeO$_6$}

\offprints{roland.mathieu@angstrom.uu.se}  

\institute{
\inst{1}\,Department of Engineering Sciences, Uppsala University, Box 534, SE-751 21 Uppsala, Sweden\\
\inst{2}\ Center of Materials Science, Department of Inorganic Materials, Karpov’ Institute of Physical Chemistry, Moscow, 105064, Russia\\
\inst{3}\,Institute for Chemical Technologies and Analytics, Vienna University of Technology, A-1060 Vienna, Austria
}

\date{Received: date / Revised version: date}

\abstract{Several $M_3$TeO$_6$ ($M$ = Mn, Co, Ni, Cu) oxides order antiferromagnetically at low temperatures ($\lesssim$ 60 K), while displaying interesting dielectric properties at high temperatures (ferroelectricity below 1000 K in $M$ = Ni case). We have investigated and analyzed the structural and magnetic properties of Mn-doped Co$_3$TeO$_6$ and Ni$_3$TeO$_6$, which order antiferromagnetically at temperatures higher than their undoped counterparts.}

\maketitle   

\section{Introduction}

There is currently a great interest in designing and finding new materials with multiferroic properties, or in a larger sense magnetic and dielectric ordering  near room-temperature. It has been found that complex crystal structures with several independent sites for magnetic cations as well as complex magnetic structures favor concomitant coupled magnetic and dielectric states\cite{mf}. 

The structural, magnetic and dielectric properties of several members of the corundum-related class of complex oxides with structural formula $A_3B$O$_6$ ($A$ cation is divalent or trivalent, $B$ cation is pentavalent or hexavalent) have been investigated, as for example the $M_3$Te$^{6+}$O$_6$ systems with $M$ = Mg$^{2+}$\cite{blasse}, Cd$^{2+}$\cite{kosse}, Mn$^{2+}$, Co$^{2+}$, Ni$^{2+}$, Cu$^{2+}$\cite{MTO,CTO,NTO,CuTO}, or ($M_2A$)Sb$^{5+}$O$_6$ compounds with $M$ = Mn$^{2+}$, Ni$^{2+}$ and $A$ = In$^{3+}$, Sc$^{3+}$\cite{MTO-InSc,NTO-InSc}. 

The $M_3$TeO$_6$ oxides with $M$ = Mn, Co, Ni, Cu order antiferromagnetically at low temperatures. Ni$_3$TeO$_6$ crystallizes in a non-centrosymmetric rhombohedral $R3$ structure, and displays a relatively simple collinear magnetic structure, consisting of ferromagnetic $ab$-planes stacked antiferromagnetically along the $c$-axis \cite{NTO}. On the contrary, the other compounds crystallize in centrosymmetric structures, with complex magnetic structures: monoclinic $C2/c$ Co$_3$TeO$_6$ displays an incommensurate, multi-propagation-vector and temperature-dependent, magnetic structure\cite{CTO}; rhombohedral $R\bar{3}$ Mn$_3$TeO$_6$ displays a two-orbit incommensurate antiferromagnetic structure\cite{MTO}; and cubic $Ia\bar{3}$ Cu$_3$TeO$_6$ orders in a ``three-dimensional spin web'' of hexagonal arrangements of magnetic moments\cite{CuTO}. 

We here report structural and magnetic properties of $M_3$TeO$_6$ single-crystals with $M$ = Mn, Co, Ni, Cu and Mn-doped $M_3$TeO$_6$ ($M$ = Co, Ni) ceramics. The antiferromagnetic interaction and antiferromagnetic transition temperature increase when Mn replaces the magnetic cation.\\

\begin{table*}[t]
 \caption{Antiferromagnetic transition temperatures $T_N$ (K) and structural parameters extracted from Rietveld refinements\cite{RIETVELD} and polyhedral analysis\cite{IVTON} for selected $M_3$TeO$_6$ samples (NPD data at 5K for $M$=Mn, Co, and Mn$_{0.2}$Co$_{0.8}$\cite{MTO,CTO,MCTO}; XRD data at 295 K for $M$=Ni and Cu\cite{NTO,CuTO} included for comparison): ionic size $r_A$ of $M^{2+}$ on $A$-site ({\AA}); space group $s. g.$; minimum, maximum, and mean values of $M$-O bond length ({\AA})  and $M$-$M$ distance ({\AA}); volume $V_A$ ({\AA}$^3$), distortion $\omega_A$, and shift from centroid $x_A$ ({\AA}) of  $M^{2+}$ in its coordination polyhedron. \textit{$^a$ $M$-O and $M$-$M$ distances were averaged over the values obtained for each non-equivalent $M(i)$ cations ($i$=5 for $M$=Co and $i$=3 for $M$=Ni); minimum-maximum values of $V_A$, $\omega_A$, and $x_A$ are indicated. $^b$ All $M^{2+}$ cations in the table have coordination number CN=6, except Co(5) which has CN=4. Smallest $V_A$ becomes 11.7 {\AA}$^3$ if only Co(1-4) with CN=6 are considered. An effective coordination CN=5 was considered to calculate Co(2)-O(i) and Co(3)-O(i) bond lengths.}}

 \begin{tabular}[htbp]{rccccccccccccc}

   & & & & $M$-O  & & & & $M$-$M$ & & & & & \\
 \cline{4-6} \cline{8-10}
$M^{2+}$ & $<$ $r_A$ $>$& $s. g.$ & min. & max.  & mean && min. & max. & mean & $V_A$ & $\omega_A$ & $x_A$ & $T_N$ \\
   \hline
    Mn  & 0.83 & $R\bar{3}$ & 2.091 & 2.373 & 2.207 && 3.21 & 3.83 & 3.56 & 12.7 & 0.103 & 0.077 & 23 \\
\hline
    Mn$_{0.2}$Co$_{0.8}$ & 0.762 & $R\bar{3}$ & 1.976 & 2.408 & 2.159 && 3.10 & 3.81 & 3.46 & 11.81 & 0.099 & 0.071 & 40 \\
\hline
$^{a,b}$Co &  0.745 & $C2/c$ & 1.856 & 2.333 & 2.064 && 2.71 & 3.93	& 3.38 & 3.5-13.0 &  0.02-0.06 & 0.06-0.24 & 26 \\
\hline
    $^a$Ni & 0.69 & $R3$ & 2.008 & 2.148 & 2.079 && 2.78 & 4.02 & 3.40 & 11.6-11.9 & 0.01-0.03 & 0.14-0.24 & 52 \\
  \hline
    Cu & 0.73 & $Ia\bar{3}$ & 1.949 & 2.369 & 2.116 && 3.18 & 3.60 & 3.39 & 10.99 &  0.105 & 0.17 & 61 \\
\hline

  \end{tabular}
  \label{table1}
\end{table*}

\section{Results and discussions}

$M_3$TeO$_6$ single crystals ($M$ = Mn, Co, Ni, Cu) and ceramics ($M$ = (Mn,Co), (Mn,Ni) were synthesized by chemical transport (see Refs.~\cite{MTO,CTO,NTO,CuTO} and \cite{MTO-MW}) and solid state reactions\cite{MTO}, respectively. The magnetic properties were investigated by magnetization and heat capacity measurements (MPMS squid magnetometer and PPMS setup from Quantum Design Inc.). Crystal and magnetic structures of the samples with $M$ = Mn, Co, and (Mn,Co) have been studied by x-ray (XRD) and neutron (NPD) powder diffraction\cite{MTO,CTO,MCTO}. Corresponding data for the samples with $M$ = Ni and Cu was obtained from literature\cite{NTO,CuTO}. The magnetic properties of ceramic samples with $M$ = Mn, Co, Ni, Cu were found to be similar to those of corresponding single crystals.

\begin{figure}[h]
\resizebox{1\columnwidth}{!}{%
  \includegraphics{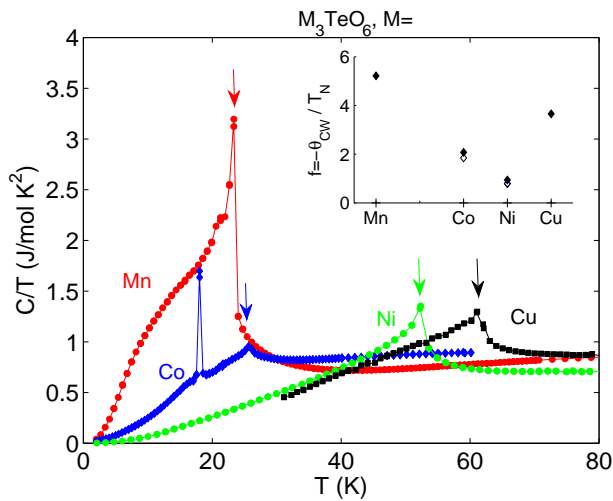}
}
\caption{Heat capacity (plotted as $C/T$) vs temperature for all single-crystals. Arrows mark the antiferromagnetic transitions. The inset shows the variation of the frustration parameter $f$ for the single crystals (filled symbols) and 20\% Mn-doped ceramics (open symbols).  
}
\label{HC}
\end{figure}

\begin{figure}[h]%
\resizebox{1\columnwidth}{!}{%
  \includegraphics{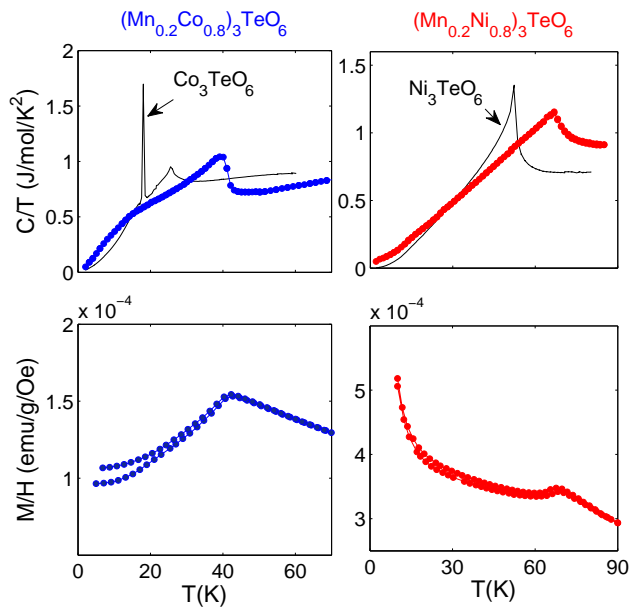}
}
\caption{
Temperature dependence of the heat capacity (upper panels, plotted as $C/T$) and magnetization recorded in $H$ = 20 Oe (lower panels, plotted as $M/H$) for 20\% Mn-doped Co$_3$TeO$_6$ and Ni$_3$TeO$_6$. $C/T$ data of undoped materials is added for comparison. 
}
\label{HCMT}
\end{figure}

Magnetic susceptibility and heat capacity experiments indicate low temperature antiferromagnetic states for all $M_3$TeO$_6$ single crystals. The magnetic transitions are evident from the heat capacity results plotted in the main frame of Fig.~\ref{HC}. The obtained antiferromagnetic transition temperatures ($T_N$), listed in Table~\ref{table1}, are in agreement with the temperatures for which $ \partial $($MT/H$)$/ \partial T$ curves is maximum. Linear fits of $H/M$($T$) data recorded up to higher temperatures were performed, yielding estimates of Curie-Weiss temperatures $\theta_{CW}$ and the magnetic frustration parameter $f$=$-\theta_{CW}/T_N$. The variation of $f$ with $M$  is illustrated in the inset of Fig.~\ref{HC}. The incommensurate magnetic structure of Mn$_3$TeO$_6$ is frustrated. For this compound, $\theta_{CW}$ $\sim$ -120 K, i.e. about 5 times $T_N$ ($f$ $\sim$ 5), to compare to $\theta_{CW}$ $\sim$ -49 K ($f$ $\sim$ 1) and $\theta_{CW}$ $\sim$  -54 K ($f$ $\sim$ 2) for Ni$_3$TeO$_6$ and Co$_3$TeO$_6$ respectively. Similarly we find $\theta_{CW}$  $\sim$ -223 K ($f$ $\sim$ 3.6) for Cu$_3$TeO$_6$. Note that in that case only the higher-temperature data ($T$ $\ge$ 200 K) closely follows a Curie-Weiss behavior (see also \cite{CuTO}). 

As seen from the heat capacity and magnetization measurements shown in Fig.~\ref{HCMT}, the antiferromagnetic ordering temperature increases in Mn-doped Co$_3$\-TeO$_6$ and Ni$_3$TeO$_6$ ceramics. In the case of (Mn$_{0.2}$Ni$_{0.8}$)$_3$TeO$_6$, the onset of antiferromagnetic ordering takes place near 67 K, i.e. above the $T_N$ of Cu$_3$TeO$_6$ (which has the highest reported undoped $T_N$ (61 K)). The $f$ parameter is nearly unaffected by Mn-doping, as shown in the inset of Fig.~\ref{HC}, reflecting the co-variation of $T_N$ and $\theta_{CW}$, and thus the increase in antiferromagnetic interaction strength.

We have also tried to dope Mn in Cu$_3$TeO$_6$; however, the temperature-dependent magnetization curves of e.g. (Mn$_{0.2}$\-Cu$_{0.8}$)$_3$TeO$_6$ show two features, near 50 K and 21 K, suggesting phase separation. Similar two phase features were obtained when doping Co in Ni$_3$TeO$_6$. For comparison, also $B$-site substitutions were attempted, e.g. Ni$_3$(Te,Mo)O$_6$, again resulting in phase separation.

\begin{figure}[h]%
\resizebox{1\columnwidth}{!}{%
  \includegraphics{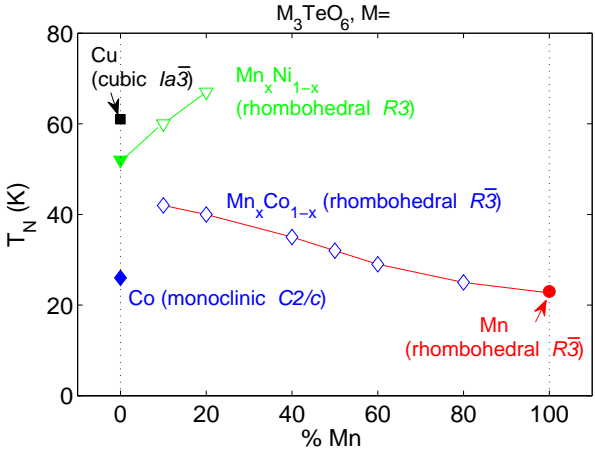}
}
\caption{
(Color online) Dependence on Mn content of the antiferromagnetic transition temperature for $M_3$TeO$_6$ single crystals (filled symbols) and solid solutions (open symbols). Connected points have the same crystal structure. 
}
\label{DIAG}
\end{figure}

The structural and magnetic properties of the pure $M_3$TeO$_6$ systems and the 20\% Mn-doped Co$_3$TeO$_6$ compound are summarized in Table~\ref{table1} and Fig.~\ref{DIAG}. The compounds contain different magnetic cations, and adopt different crystal structures with sometimes non-equivalent crystallographic and magnetic sites. It is thus difficult to determine an exact correlation between structural and magnetic properties. We have attempted to extract more structural parameters from the refinements, such as $M$-O-$M$ bond angles, but those angles were found to vary greatly (e.g. between 29$^o$  and 161$^o$ for $M$ = Mn). Nevertheless, as illustrated in Table~\ref{table1} and Fig.~\ref{DIAG}, while Mn-doped Ni$_3$TeO$_6$ remains rhombohedral $R3$, Mn-doped Co$_3$\-TeO$_6$ surprisingly adopts the rhombohedral $R\bar{3}$ structure of Mn$_3$TeO$_6$ (even with only 10\% Mn, see also Ref.~\cite{MCTO}). As seen in the table, all the obtained structural and polyhedral parameters for Mn$_3$TeO$_6$ and (Mn$_{0.2}$Co$_{0.8}$)$_3$TeO$_6$ are relatively similar.  Interestingly, by extrapolating the data presented in Fig.~\ref{DIAG}, one can predict that if rhombohedral $R\bar{3}$ Co$_3$TeO$_6$ could be stabilized (e.g. by synthesis under pressure), it would have a $T_N$ of about 45 K, i.e. nearly twice that of monoclinic Co$_3$TeO$_6$. Similarly, assuming that Mn-doped Ni$_3$TeO$_6$ remains in the rhombohedral $R3$ structure for large amounts of Mn, such an Mn-rich (Ni,Mn)$_3$TeO$_6$ or pure Mn$_3$TeO$_6$ could order antiferromagnetically at temperatures around 100 K. 

Dielectric and ferroelectric properties of some $M_3$TeO$_6$ ($M$= Cd, Mg) have been investigated in earlier studies\cite{blasse,kosse}. More recently, it was observed that Co$_3$TeO$_6$ could acquire an electronic polarization at low temperatures\cite{CTO-POL}, and that Ni$_3$TeO$_6$ is ferroelectric below 1000 K\cite{NTO-InSc}. Since the non-centrosymmetric structure of Ni$_3$TeO$_6$ is preserved upon doping with Mn, our results suggest that rhombohedral $R3$ Mn-rich (Ni,Mn)$_3$TeO$_6$  or pure Mn$_3$TeO$_6$ would display enhanced magnetic properties (higher $T_N$) while retaining ferroelectric properties. 

\section{Conclusions}

We have observed significant changes in the structural and magnetic properties of $M_3$TeO$_6$ solid solutions. For example, the replacement of one fifth of the magnetic cation with Mn in Ni$_3$TeO$_6$ and Co$_3$TeO$_6$ yields an increase in antiferromagnetic interaction and transition temperatures. We predict that new $(M',M'')_3$TeO$_6$ multiferroics with higher magnetic ordering temperature, such as rhombohedral $R3$ Mn-rich (Ni,Mn)$_3$TeO$_6$ could be designed and synthesized.

\section {acknowledgement}
We thank the Swedish Research Council (VR), the G\"oran Gustafsson Foundation, and the Russian Foundation for Basic Research for financial support.

\end{document}